\documentclass[%
reprint,
superscriptaddress,
 amsmath,amssymb,
 aps,
]{revtex4-1}

\usepackage{graphicx}
\usepackage{dcolumn}
\usepackage{bm}

\begin{document}

\preprint{Draft}

\title{Disorder induced helical-edge transport near $\nu$=0 of monolayer graphene}

\author{Sung Ju Hong}
\affiliation{Institut f\"u{}r Festk\"o{}rperphysik, Leibniz Universit\"a{}t Hannover, 30167 Hannover, Germany}
\affiliation{Present address: Department of Energy Science, Sungkyunkwan University, Suwon, 16419, Republic of Korea}
\author{Christopher Belke}
\affiliation{Institut f\"u{}r Festk\"o{}rperphysik, Leibniz Universit\"a{}t Hannover, 30167 Hannover, Germany}
\author{Johannes C. Rode}
\affiliation{Institut f\"u{}r Festk\"o{}rperphysik, Leibniz Universit\"a{}t Hannover, 30167 Hannover, Germany}
\author{Benedikt Brechtken}
\affiliation{Institut f\"u{}r Festk\"o{}rperphysik, Leibniz Universit\"a{}t Hannover, 30167 Hannover, Germany}
\author{Rolf J. Haug}
\affiliation{Institut f\"u{}r Festk\"o{}rperphysik, Leibniz Universit\"a{}t Hannover, 30167 Hannover, Germany}

\date{\today}

\begin{abstract}
The complex nature of filling factor $\nu$=0 of monolayer graphene is studied in magnetotransport experiments. As function of perpendicular magnetic field a metal-insulator transition is observed, which is attributed to disorder-induced Landau level broadening in the canted antiferromagnetic phase. In the metallic regime a separation of the zeroth Landau level appears and signs of the quantum spin Hall effect are seen near $\nu$=0.  In addition to local transport, nonlocal transport experiments show results being consistent with helical edge transport. 

\end{abstract}

\pacs{Valid PACS appear here}
\maketitle


The zeroth Landau level (zLL) of monolayer graphene (MLG) shows a remarkable quantum Hall behaviour because the zLL consists of both electron and hole-like dispersions near the edge \cite{Nature.438.197,Nature.438.201,PhysRevLett.95.146801,PhysRevB.73.195408}. Furthermore, half filling of the zLL ($\nu$=0) can exhibit quite a complex behaviour with metallic and insulating phases being attributed to spin-polarized \cite{PhysRevLett.96.176803,PhysRevLett.97.116805,PhysRevLett.98.196806} and valley-polarized \cite{PhysRevLett.96.136806,PhysRevLett.98.016803,PhysRevLett.100.176404,PhysRevLett.108.106804} states, respectively. In order to unveil the nature of the filling factor $\nu$=0, intensive studies have been carried out \cite{PhysRevLett.96.176803,PhysRevLett.97.116805,PhysRevLett.98.196806,PhysRevLett.96.136806,PhysRevLett.98.016803,PhysRevLett.100.176404,PhysRevLett.108.106804,PhysRevLett.96.256602,PhysRevLett.99.106802,PhysRevLett.100.206801,PhysRevB.79.115434,PhysRevB.80.201403,PhysRevB.80.235417,PhysRevLett.105.046804,NewJPhys.13.113008,PhysRevB.85.155439,PhysRevB.86.075450,NaturePhys.8.550}. Especially, quantum Hall ferromagnetism (QHFM) described by SU(4) isospin has provided a theoretical framework to elucidate the partial filling of the zLL \cite{PhysRevLett.108.106804,PhysRevLett.96.256602,NaturePhys.8.550,PhysRevB.92.201412} and various quantum phases of $\nu$=0 \cite{PhysRevB.85.155439,PhysRevB.86.075450}. As a result, phases for a metallic state (ferromagnetic (FM)) and for insulating states (charge-density-wave (CDW), Kekul\'e{} distortion (KD), antiferromagnetic (AF), and canted antiferromagnetic (CAF)) have been suggested \cite{PhysRevB.85.155439}. 

With the availability of high quality devices, a spin-unpolarized insulating state \cite{PhysRevLett.108.106804,NaturePhys.8.550,PhysRevB.92.201412} was observed and the nature of the insulating state was revealed to be the CAF phase \cite{Nature.505.528}. This work showed the transition from the CAF to FM phase by the application of an additional parallel magnetic field, which triggered researches related to the different quantum phases of $\nu$=0 \cite{PhysRevB.90.195407,PhysRevB.90.241410,PhysRevB.92.165110,PhysRevB.93.045105,PhysRevB.93.115137,PhysRevB.94.085135,PhysRevB.94.245435}. Subsequently, spin superfluidity and magnon generation/detection have been investigated, based on the CAF phase at $\nu$=0 and applications in terms of spin transport have been put forward \cite{PhysRevLett.116.216801,NaturePhys.14.907,Science.362.229}. As the CAF phase changes to the FM phase in strong parallel magnetic fields, quantum spin Hall (QSH) effect is obtained due to the formation of helical edge states. However, the helical edge structure can be obtained even in the CAF phase, where counter-propagating edge modes are formed with oppositely canted antiferromagnetic spin textures \cite{PhysRevB.86.075450,Nature.505.528}. Therefore, novel transport phenomena due to the helical band structure are expected in the CAF phase.

\begin{figure*}[]
\begin{center}
\includegraphics[width=2\columnwidth]{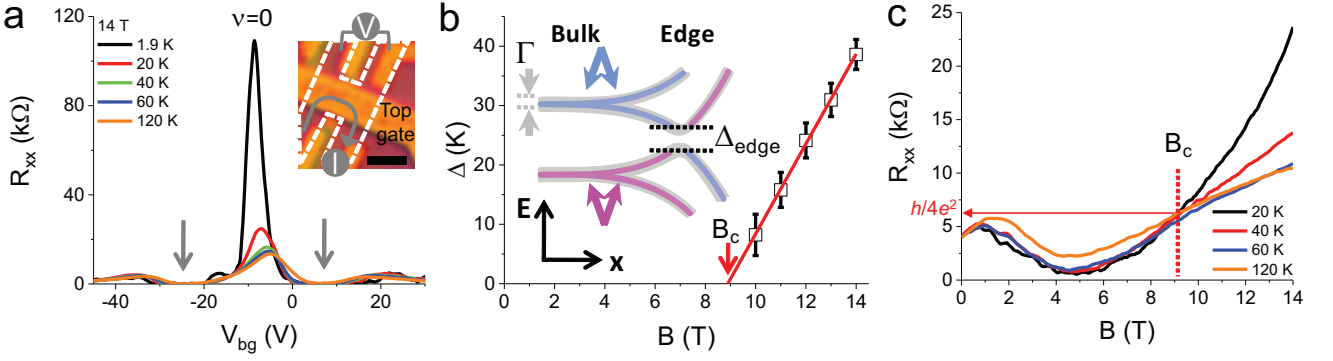}

\end{center}
\vspace*{-0.6cm}
\caption{(color) (a) Temperature dependent $R$$_{xx}$ at 14 T. Gray arrows indicate $\nu$=$\pm$2. Inset shows an optical microscope image and the white dashed H-bar indicates MLG device. $I$ and $V$ are bias current and voltage drop, respectively. Scale bar is 2 $\mu$m. Inset of (b) is schematic band structure of the CAF phase, where $\Gamma$ and $\Delta_{edge}$ indicate Landau level broadening and edge gap, respectively. (b) $\Delta$ vs $B$ at 1.9 K. The red line is a fitted line, which yields $B$$_{c}$$\approx$8.9 T and $\Gamma$$\approx$67 K. (c) $R$$_{xx}$ vs $B$ near $\nu$=0 for different temperatures.}
\end{figure*}

Here, we investigate helical edge transport of MLG at $\nu$=0 in terms of local and nonlocal measurements. We found a metal-insulator transition (MIT) at a critical perpendicular magnetic field $B$$_{c}$ which can be attributed to a disorder-induced broadening of the CAF state. Above $B$$_{c}$ typical transport features confirm the CAF phase. Below $B$$_{c}$ a splitting of the zLL and conductance saturation are observed, which interestingly indicate the formation of helical edge states also in this phase. We performed additional parallel magnetic field experiments, which identified the nature of the helical edge states. Furthermore, nonlocal transport experiments support the result of the local transport measurements.

Magnetotransport was investigated in a H-bar geometry of MLG encapsulated by hexagonal boron nitride (hBN) \cite{Science.342.614} as shown in the inset of Fig. 1(a). The MLG was obtained by mechanical exfoliation of natural graphite and identified with red-filtered optical contrast on SiO$_{2}$(330 nm)/Si substrate. The encapsulation by hBN(HQ Graphene) was carried out by pick up and transfer method \cite{Science.342.614}. Top (13 nm thick) and bottom (120 nm thick) hBN were intentionally misaligned to minimize the perdiodic modulation of the coupling between MLG and the hBNs. The device was annealed with forming gas (5 $\%$ H$_{2}$ and 95 $\%$ N$_{2}$) at 350 $^{\circ}$C and subsequently, atomic force microscope (AFM) measurements were performed to select a bubble-free region. Metallic edge-contacts (Cr/Au=8/62 nm) were evaporated after a SF$_{6}$ reactive ion etching process. Electrical measurements were performed by DC (10 nA excitation current) method in the variable temperature insert of a superconducting magnet system with magnetic fields up to 14 T. The top gate voltage was fixed to be ground during the measurements.

Figure 1(a) shows the longitudinal resistance $R$$_{xx}$ as function of bottom gate voltage ($V$$_{bg}$) for different temperatures at 14 T, indicating an insulating state at $\nu$=0. From $R_{xx}$$\propto$$exp(\Delta/2T)$, we extracted a thermal activation gap, $\Delta$, for $\nu$=0 as a function of magnetic field as shown in Fig. 1(b). The activation gap increases linearly above $B$$_{c}$, which is in agreement with previous observations \cite{NaturePhys.8.550}. It is worth to note while an interaction-induced gap usually shows a square root dependence with magnetic field \cite{PhysRevB.88.115407}, sub-linear and linear dependences are also possible in the case of finite-range Coulomb interaction \cite{PhysRevB.90.201409}. The nature of the observed insulating state is known as the CAF phase and the corresponding schematic band diagram is seen in the inset of Fig. 1(b). From the intercept at $B$=0 T of the magnetic field dependent thermal activation gap, we extract a level broadening of $\Gamma$$\approx 67~$K which we attribute to disorder. The valley isospin anisotropy energy $u_{\bot}$$\approx$1-10$B$[T] K \cite{PhysRevB.86.075450} governing the gap in the CAF at the edge is of the same order of magnitude as $\Gamma$$\approx$67 K for a magnetic field of $B$$_{c}$$\approx$8.9 T. The disorder induces a level broadening (thicker gray curve in the schematic band diagram) and as a result the gap at the edge is effectively closed ($\Gamma$$>$$\Delta_{edge}$).

Figure 1(c) shows $R$$_{xx}$ along $\nu$=0 as a function of magnetic field for different temperatures. At $B$$_{c}$$\approx$8.9 T, a MIT is observed with a $B$$_{c}$ value being consistent with the thermal activation gap analysis in Fig. 1(b).  The metallic state observed below $B$$_{c}$ has to be attributed to the disappearance of the energy gap near the edge. This is in agreement with general consensus that the insulating state tends to be observed in cleaner devices ($\Gamma$$<$$\Delta_{edge}$) \cite{PhysRevB.79.115434,PhysRevB.80.235417} at lower magnetic fields. In Fig. 1 (c) a clear minimum in $R$$_{xx}$ is observed around 5 T indicating edge-state transport as in usual quantum Hall systems. It is noteworthy that the $R$$_{xx}$ value at $B$$_{c}$ is around $h$/4$e$$^{2}$. This value can be attributed to full equilibration of counter-propagating edge states in metal electrodes \cite{NatureNanotechnol.12.118}.

Now we discuss further this metallic state below $B$$_{c}$. Figure 2(a) shows $R$$_{xx}$ vs $V$$_{bg}$ at 4~T for different temperatures. We observe a splitting of the zLL (zLL$_{-}$ and zLL$_{+}$ depicted by red squares). Furthermore, we confirm the metallic state by two-terminal conductance ($G$) measurements. Figure 2(b) shows the temperature dependence of $G$ at 5 T, where $G$ exhibits metallic behaviour near $\nu$=0. Note that we corrected for the contact resistance ($R_{c}$$\approx$0.3 k$\Omega$) by matching the expected quantization at $\nu$=-2. The used measurement configuration is shown in the inset of Fig. 2(b). The conductance $G$ measured in this configuraten can be seen as function of magnetic field for temperatures varied between 60 K and 20 K in Fig. 2(c). Interestingly, in a broad range of intermediate magnetic fields, $G$ shows a saturation behaviour with a value of 1.3$e$$^{2}$/$h$ (orange bar shown in Fig. 2(c)). This conductance saturation close to the ideal value, 2$e$$^{2}$/$h$, seems to evidence a helical quantum spin Hall (QSH) state. A similar conductance saturation was previously observed in graphene electron-hole bilayers, where filling factors $\nu$=$\pm1$ are existing at the same time in the two different layers giving a total filling factor of $\nu$=0 \cite{NatureNanotechnol.12.118}. In our system the helical edge states are formed at $\nu$=0 due to the specific edge structure of a monolayer of graphene where the exchange interaction at the edge is suppressed by disorder (see inset in Fig. 2(c)). As a result, the conductance saturation observed in our system can be regarded as an indication of QSH effect. The deviation of the ideal quantization value can be attributed to backscattering and is in agreement with the theoretical expectations for finite temperatures \cite{PhysRevB.93.115137}. Likewise in $R$$_{xx}$, the transition to the insulating state is observed for $G$ at the same $B$$_{c}$ (red arrow in Fig. 2(c)). According to Young $et$ $al$.,\cite{Nature.505.528} the $G$ corresponding to QSH state is determined by the measurement configuration taking equilibration of edge states in contacts into account (see e.g. \cite{Semicond.1993}). As a result, the two-terminal conductance depends on the number of floating contacts along each edge between source and drain contact and the conductance reads $G$=$e$$^{2}$/$h$((N$_{1}$+1)$^{-1}$+(N$_{2}$+1)$^{-1}$) with N$_{1}$ and N$_{2}$  the number of floating metal electrodes along the two edges. To test this Fig. 2(d) shows $G$ measured with the configuration 2 sketched in the inset where on both edges a floating contact exists. A clear saturation with a value close to $e$$^{2}$/$h$ (depicted by orange bar in Fig. 2(d)) is observed for an intermediate magnetic field range being perfectly consistent with the expectation according to the above formula, i.e. helical edge transport.  The relatively large deviation from the expected value of 2$e$$^{2}$/$h$ in case of configuration 1 (Fig. 2(c)) is attributed to the presence of additional backscattering.

\begin{figure}[]
\begin{center}
\includegraphics[width=\columnwidth]{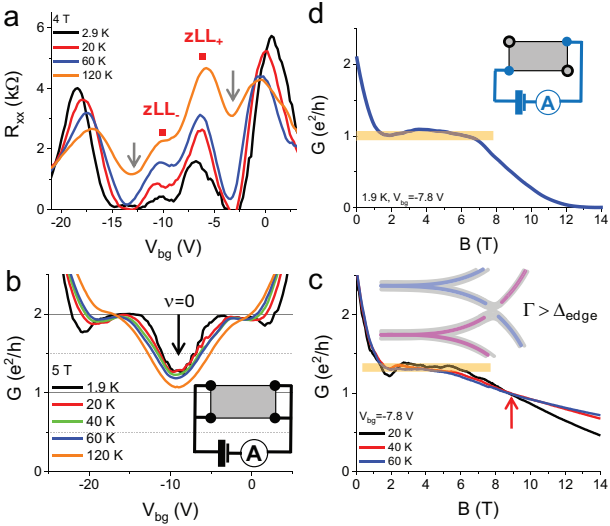}

\end{center}
\vspace*{-0.6cm}
\caption{(color) (a) $R$$_{xx}$ vs $V$$_{bg}$ at 4 T for different temperatures. (b) $G$ vs $V$$_{bg}$ at 5 T for different temperatures. Inset indicates measurement configuration of $G$ for (b) and (c). (c) $G$ vs $B$ near $\nu$=0 for different temperatures. Inset indicates schematic band diagram of the CAF phase where the gap is effectively closed due to disorder-induced broadening. (d) $G$ vs $B$ with configuration 2 depicted by inset.}
\end{figure}

The nature of the helical edge states at $\nu$=0 is influenced  by parallel magnetic fields. It has been reported that the CAF phase is changed to the FM phase by application of strong parallel magnetic fields \cite{Nature.505.528}.  We performed also parallel field experiments with tilting our graphene device (see inset of Fig. 3(a)). Note that the parallel magnetic field has both parallel and perpendicular components compared to the current direction. Tilting angles were estimated from the Landau fan diagram which is determined by the perpendicular component ($B_{\bot}$) of the applied external magnetic field ($B_{ext}$). Figure 3(a) shows $R$$_{xx}$ versus $B_{\bot}$. We explored three different tilting cases such as 90$^{\circ}$ ($B_{\bot}$=$B_{ext}$), 56.1$^{\circ}$, and 41.6$^{\circ}$, respectively. We observed that $R$$_{xx}$ increases at fixed perpendicular magnetic field with additional parallel ($B_{\parallel}$) magnetic field, while previous observations for the insulating CAF phase exhibited a decreasing $R$$_{xx}$ with additional parallel magnetic field \cite{NaturePhys.8.550}. We attribute this different behaviour to the small ratio between parallel and perpendicular magnetic field used here in comparison to previous works where extremely large parallel fields were applied. We also carried out temperature dependent experiments for all cases, which reveal a similar $B$$_{c}$ for all tilt angles (as seen in the inset of Fig. 3(b)). The invariance of $B$$_{c}$ with parallel field indicates that the effective gap opening leading to the insulating state is determined only by the perpendicular magnetic field. In other words, perpendicular magnetic field seems to determine the band structure and the parallel field additionally affects transport in our case. 

\begin{figure}[]
\begin{center}
\includegraphics[width=\columnwidth]{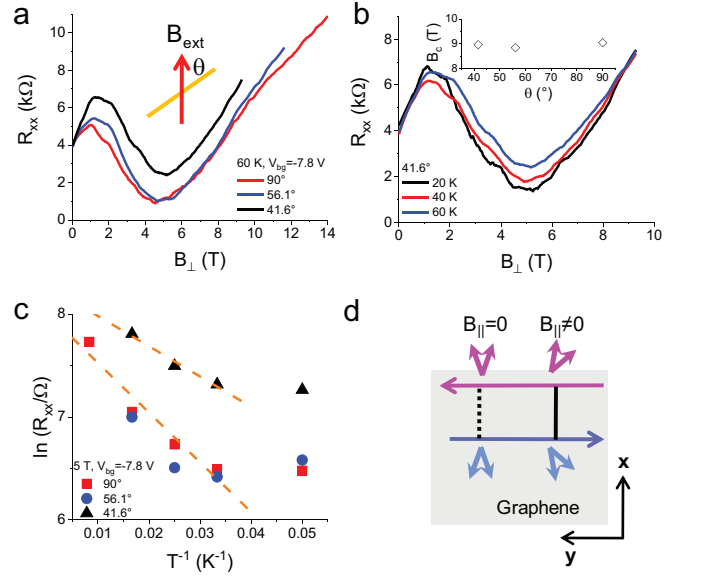}

\end{center}
\vspace*{-0.6cm}
\caption{(color) (a) Parallel-magnetic-field dependent $R$$_{xx}$ vs $B_{\bot}$ near $\nu$=0 at 60 K. Inset indicates tilting angle ($\theta$) between external magnetic field ($B_{ext}$) and graphene plane (depicted by yellow line) (b) Temperature dependent $R$$_{xx}$ vs $B_{\bot}$ for 41.6$^{\circ}$ tilted case. Inset indicates $B$$_{c}$ in terms of perpendicular component. (c) Arrhenius plot of $R$$_{xx}$ at 5 T perpendicular field. The orange dashed lines are fitted and yield thermal activation energies. (d) Schematic diagram of helical edge channel and backscattering with (black solid line) and without (black dotted line) parallel magnetic field.}
\end{figure}

Figure 3(b) shows $R$$_{xx}$ for different temperatures in the case of 41.6$^{\circ}$ as function of perpendicular magnetic field. It indicates metallic edge-state transport below $B$$_{c}$ as in the 90$^{\circ}$ case. Analysing the temperature dependence of the minimum observed at $B_{\bot}$=5~T we obtain an activated behaviour as in quantum Hall systems.  Figure 3(c) shows Arrhenius plots of $R$$_{xx}$ at $B_{\bot}$=5~T, yielding thermal activation gaps of 59 K at 41.6$^{\circ}$ and 97 K at 90$^{\circ}$. First of all, the thermal activation gap becomes smaller for additionally applied parallel magnetic field. Furthermore, the parallel field does not seem to affect transport in a linear fashion. That is, the 56.1$^{\circ}$ tilted case is similar with the 90$^{\circ}$ tilted case (within our error bars) and a significant reduction in activation energy is only observed for the 41.6$^{\circ}$ tilted case. Therefore, we can exclude the possibility of Zeeman splitting where the gap should become gradually larger with larger total magnetic field. In contrast we observe a reduction of the gap. The measured gap corresponds to the energy governing the disordered CAF phase. The increased $R$$_{xx}$ at low temperatures seems to result from increased backscattering between the helical edge states. In the absence of parallel magnetic field, the backscattering depicted by black dotted line in Fig. 3(d) is suppressed because the two edge states have oppositely canted antiferromagnetic spin textures. Application of a parallel magnetic field seems to enhance the disorder-induced scattering between the helical edge states, which leads to a stronger backscattering (black solid line in Fig. 3(d)). As a result, $R$$_{xx}$ increases.

\begin{figure}[]
\begin{center}
\includegraphics[width=\columnwidth]{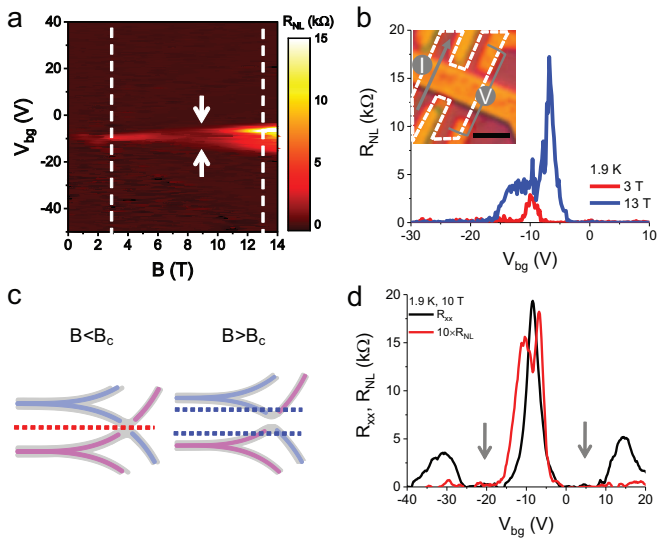}

\end{center}
\vspace*{-0.6cm}
\caption{(color) (a) Mapping of $R$$_{NL}$ as functions of $V$$_{bg}$ and $B$. (b) Line-profiles of $R$$_{NL}$ at 3 T (red) and 13 T (blue), respectively. Inset is measurement configuration of $R$$_{NL}$. Scale bar is 2 $\mu$m. (c) Schematic diagram of band structure for 3 T (left) and 13 T (right), respectively. Red and blue dotted lines are Fermi level where the helical transport dominates. (d) $R$$_{xx}$ (black) and $R$$_{NL}$ (red) at 10 T. Gray arrows indicate $\nu$=$\pm$2.}
\end{figure}

To complement the local transport measurements, we measured nonlocal resistance, $R$$_{NL}$ as function of carrier density and perpendicular magnetic field at 1.9 K (shown in Fig. 4(a)). At a perpendicular magnetic field of 3 T (white dashed line) in Fig. 4(a), a single peak is observed at $\nu$=0 (red curve shown in Fig. 4(b)). At 13 T, the $R$$_{NL}$ profile (second white dashed line in Fig. 4(a)) has evolved into a double peak structure as shown by the blue curve of Fig. 4(b). Several origins such as classical Ohmic contribution, flavor Hall effect (FHE) \cite{Science.332.328}, and helical QSH effect \cite{NatureNanotechnol.12.118,NatureCommun.8.2198} can yield finite values of $R$$_{NL}$ near the Dirac point. First of all, we rule out the classical Ohmic contribution which is proportional to $R$$_{xx}$. As a counter example of the Ohmic contribution, $R$$_{xx}$ and $R$$_{NL}$ at 10 T do not show a proportional behaviour as shown in Fig. 4(d). The FHE does also not seem to be related with our result in terms of $R$$_{NL}$ in the insulating state. While a single peak is obtained from the FHE at $\nu$=0 \cite{Science.332.328} in the insulating state, we observed a local minimum at $\nu$=0 in the insulating state and a double peak structure at $\nu$$\neq$0. In contrast, the helical QSH effect can provide a consistent explanation. $R$$_{NL}$ has finite values in the helical QSH state \cite{Science325.294}. In the metallic regime, the helical edge state is dominating at $\nu$=0 (left panel of Fig. 4(c)), which yields a single peak of $R$$_{NL}$. The single peak of $R$$_{NL}$ due to the helical QSH effect is in agreement with the results in the electron-hole bilayer \cite{NatureNanotechnol.12.118} mentioned before and experiments studying grain-boundary scattering in polycrystalline graphene \cite{NatureCommun.8.2198}. Near $B$$_{c}$, $R$$_{NL}$ starts to split as shown by the white arrows in Fig. 4(a) due to the gap opening. In the insulating regime, the helical edge states disappear at the gap but survive for both carrier density regimes slightly away from $\nu$=0 (right panel of Fig. 4(c)), which yields the double peaks. Therefore, the evolution of $R$$_{NL}$ is also supporting the helical transport property near $\nu$=0 as discussed for the local transport results.

Combining helical edge channels and superconductors \cite{NaturePhys.14.411} can give origin to majorana zero modes which are extremely interesting for applications in quantum computing. In the case of graphene, helical edge channels were only expected for the FM phase since the intrinsic Zeeman energy is lower than the valley isospin anisotropy energy. Therefore, different approaches were used to generate the FM phase necessary for helical transport at the edge. As discussed above applying extremely large parallel magnetic fields generates the FM phase in graphene. Other approaches include the use of different materials in close proximity to MLG. So, e.g. magnetic materials can be brought in proximity to graphene to promote the FM phase at $\nu$=0 \cite{NatureMater.15.711,NanoLett.18.2435,PhysRevB.95.195426,arXiv:1905.06866}. In another approach screening effects by high-$k$ dielectric substrate, SrTiO$_{3}$ are used to produce the FM phase due to the suppression of Coulomb interaction \cite{arXiv:1907.02299}. Compared to these phase-engineering approaches our result here implies that helical transport is feasible even in pristine graphene. Namely, the edge gap in the CAF phase can be effectively closed by disorder-induced broadening so that helical transport can be obtained. This result where one needs a certain strength of disorder to have helical edge transport in an intermediate magnetic field regime may provide a simpler way to obtain helical transport in comparison to applying additional large parallel magnetic fields or in comparison to employing different host materials in close proximity to MLG.  

In conclusion, we have studied magnetotransport properties of MLG at $\nu$=0. We have observed a metal-insulator transition at a critical perpendicular magnetic field of $B$$_{c}$= 8.9 T, which is  explained by the influence of disorder-induced broadening in the CAF phase. Below $B$$_{c}$, helical transport is dominating at $\nu$=0, which is confirmed by the observed splitting of the zLL and conductance saturation. Furthermore, experiments as function of parallel magnetic field for the metallic phase reveal a reduction of the relevant gap and that spin-related scattering between the helical edge states is involved. Finally, we obtained consistent transport results in terms of nonlocal measurement.

\begin{acknowledgments}
The authors gratefully acknowledge valuable discussions with P. G. Silvestrov and P. Recher. This work is supported by School for Contacts in Nanosystems, DFG within priority program SPP 1459, Joint Lower Saxony-Israeli research project, and Fundamentals of Physics and Metrology Initiative of Lower Saxony. Part of this study has been performed using facilities at the LNQE, Leibniz Universit\"a{}t Hannover, Germany.
\end{acknowledgments}

\nocite{*}

\bibliography{Draft}

\end{document}